\documentclass[conference,a4paper]{APSIPA2021}
\usepackage{amsmath}
\usepackage{amsfonts}
\usepackage{graphicx}
\usepackage{multirow}
\usepackage{booktabs}
\usepackage{threeparttable}
\usepackage{amssymb}
\usepackage{float}
\usepackage[backend=biber,style=ieee,]{biblatex}
\usepackage{bm}
\usepackage{geometry}
\usepackage{fancyhdr}

\fancypagestyle{firststyle}{
  \fancyhf{}
  \fancyhead[C]{2025 Asia Pacific Signal and Information Processing Association Annual Summit and Conference (APSIPA ASC)}
}

\begin{document}

\title{DARS: Dysarthria-Aware Rhythm-Style Synthesis for ASR Enhancement}

\author{
\authorblockN{
Minghui Wu\authorrefmark{1}, Xueling Liu\authorrefmark{2}, Jiahuan Fan\authorrefmark{2}, Haitao Tang\authorrefmark{2}, Yanyong Zhang\authorrefmark{1}, Yue Zhang\authorrefmark{3}
}

\authorblockA{
\authorrefmark{1}
University of Science and Technology of China, China
\\
E-mail: wmhsky@mail.ustc.edu.cn, yanyongz@ustc.edu.cn
}

\authorblockA{
\authorrefmark{2}
iFlytek Co., Ltd., China}

\authorblockA{
\authorrefmark{3}
Huawei Technology, China}


}

\maketitle
\thispagestyle{firststyle}
\pagestyle{fancy}

\begin{abstract}
Dysarthric speech exhibits abnormal prosody and significant speaker variability, presenting persistent challenges for automatic speech recognition (ASR). While text-to-speech (TTS)-based data augmentation has shown potential, existing methods often fail to accurately model the pathological rhythm and acoustic style of dysarthric speech. 
To address this, we propose DARS, a dysarthria-aware rhythm-style synthesis framework based on the Matcha-TTS architecture. DARS incorporates a multi-stage rhythm predictor optimized by contrastive preferences between normal and dysarthric speech, along with a dysarthric-style conditional flow matching mechanism, jointly enhancing temporal rhythm reconstruction and pathological acoustic style simulation. Experiments on the TORGO dataset demonstrate that DARS achieves a Mean Cepstral Distortion (MCD) of 4.29, closely approximating real dysarthric speech.
Adapting a Whisper-based ASR system with synthetic dysarthric speech from DARS achieves a 54.22\% relative reduction in word error rate (WER) compared to state-of-the-art methods, demonstrating the framework's effectiveness in enhancing recognition performance.
\end{abstract}

\section{Introduction}
Dysarthria is a motor speech disorder caused by neurological conditions and is typically characterized by slurred articulation, reduced speech rate, and abnormal prosody.
These manifestations significantly impair verbal communication and limit the social participation of dysarthric patients~\cite{duffy2012motor, enderby2013disorders}. 
Automatic speech recognition (ASR), as an assistive tool, holds promise for assisting individuals with dysarthria~\cite{vachhani2018data, qian2023survey}.
However, current ASR systems still exhibit limited recognition accuracy for dysarthric speech. The main challenges arise from two key factors: (1) substantial speaker variability stemming from heterogeneous dysarthric patterns~\cite{mustafa2015exploring, rowe2022characterizing}, and (2) severe data scarcity due to the high cost and complexity of data collection. These issues hinder the development of robust ASR systems tailored to dysarthric speech~\cite{young2010difficulties, bhat2025speech}.

To address data scarcity, recent studies have explored text-to-speech (TTS)-based data augmentation to reduce the notable prosody mismatch between synthesized and real dysarthric speech~\cite{soleymanpour2022synthesizing, wagner2025personalized, leung2024training, el2025unsupervised}, particularly in terms of rhythm and style distribution. Wagner \textit{et al.}~\cite{wagner2025personalized} integrate large language models with controllable TTS and x-vector-based speaker adaptation, achieving performance gains for severe dysarthria. However, the method relies heavily on prompt engineering and offers limited controllability over speech style. Leung et al.~\cite{leung2024training} use a diffusion-based approach to synthesize pathological speech with high-speech-quality and semantic consistency, but lack precision in modeling dysarthric rhythm patterns. Although Soleymanpour \textit{et al.}~\cite{soleymanpour2022synthesizing} introduce severity-conditioned style controller and pause insertion mechanism, their approach remains limited in its ability to capture subtle speech characteristics in mild dysarthric cases.

To address these limitations in rhythm modeling precision and pathological style controllability, we propose the \textbf{D}ysarthria-\textbf{A}ware \textbf{R}hythm-\textbf{S}tyle (DARS) synthesis framework, based on the Matcha-TTS architecture. DARS comprises two key mechanisms: 
\begin{enumerate}
    \item A \textbf{multi-stage rhythm predictor} that follows a pause-then-duration strategy, guided by contrastive preference optimization (CPO) between normal and dysarthric speech to better capture fragmented rhythmic patterns;
    \item A \textbf{dysarthria-aware conditional flow matching} mechanism that incorporates pathological style vectors to constrain the synthesis process, thereby enhancing modeling of dysarthric acoustic variations.
\end{enumerate}
These two mechanisms work synergistically to significantly improve the prosody consistency of synthesized dysarthric speech, generating highly representative training data for dysarthric ASR systems.

\section{RELATED WORKS}

\begin{figure*}[t]
 \centering
 \centerline{\includegraphics[width=0.8\textwidth]{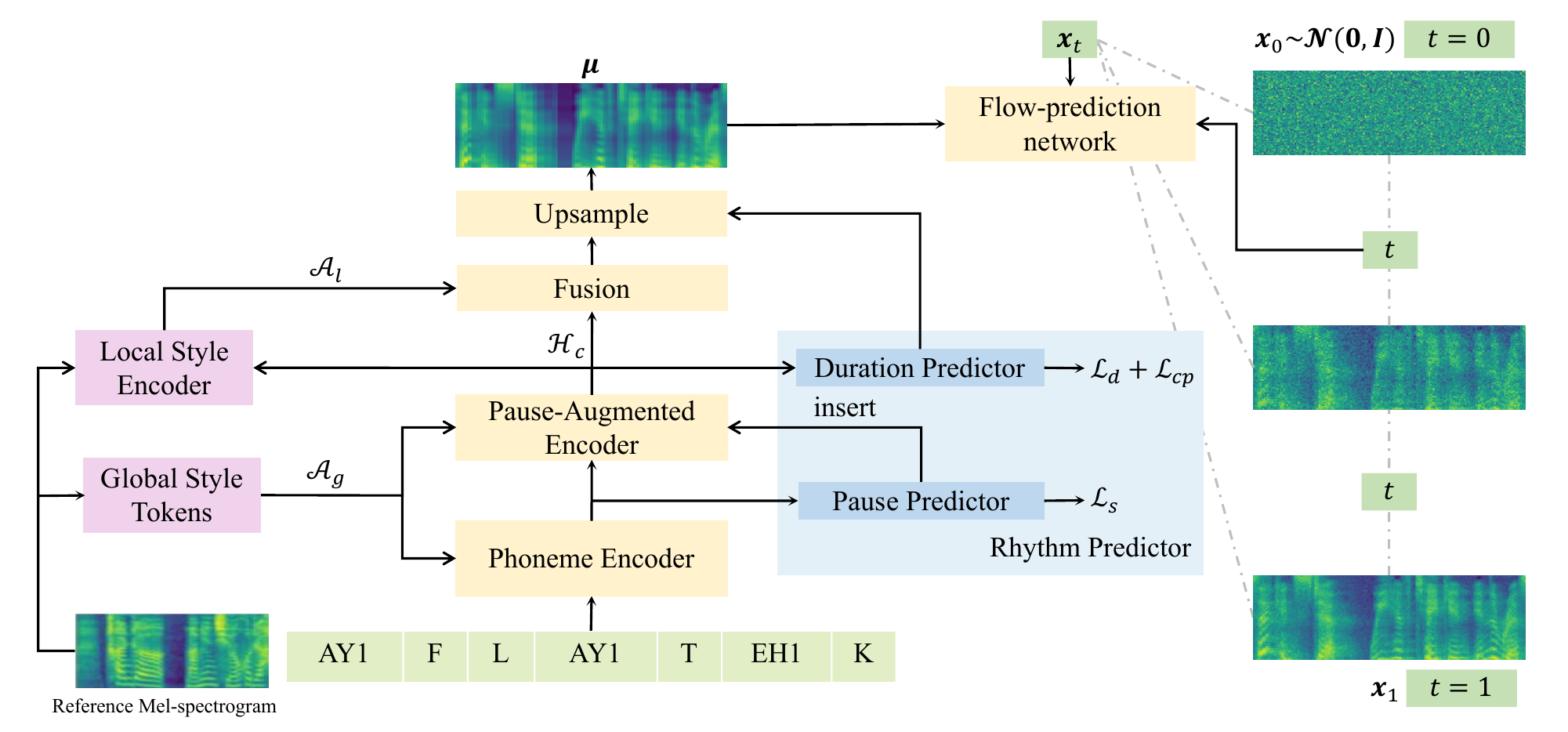}}
\caption{DARS framework based on Matcha-TTS model}
\label{framework_v1}
\end{figure*}

\subsection{MATCHA-TTS}
Given the computationally intensive sampling process in diffusion probabilistic model (DPM)-based TTS systems~\cite{popov2021grad, huang2022prodiff}, we adopt Matcha-TTS~\cite{mehta2024matcha} as our backbone. This non-autoregressive encoder-decoder (ED) framework enables high-fidelity speech synthesis with significantly reduced inference steps, while achieving automatic alignment between input text and dysarthric speech without manual annotation.

The encoder employs a transformer-based architecture~\cite{vaswani2017attention} to extract high-level semantic representations from text. The monotonic alignment search (MAS)~\cite{kim2020glow} provides forced alignment information as targets for the duration predictor. During inference, the duration predictor's outputs are used to upsample the encoder outputs, yielding the conditional mean values $\mathbf{\mu}$. The decoder adopts U-Net-based flow-prediction network~\cite{rombach2022high}, consisting of downsampling, middle and upsampling modules. Each module is composed of transformer layers, and the Snake-beta activation from BigVGAN~\cite{lee2022bigvgan} is used to accelerate convergence.

During training, Matcha-TTS model eliminates the redundant multi-step denoising process in DPMs by employing an Optimal-Transport Conditional Flow Matching (OT-CFM) strategy~\cite{huang2022prodiff, ye2023comospeech, mehta2024matcha}. This method simplifies the mapping from random noise to target speech by modeling the conditional flow-matching vector field. Acoustic features are generated by solving the corresponding ordinary differential equation (ODE) using a first-order Euler forward method~\cite{song2020score, albergo2022building}, reducing more than 50\% inference steps.

\subsection{Whisper-Based ASR}
Whisper is one of the most widely adopted end-to-end speech recognition frameworks~\cite{cao2012whisper}, featuring a transformer-based ED architecture. The encoder receives mel-spectrograms as input and produces high-level semantic representations, while the decoder generates the corresponding text sequence in an autoregressive manner. The model is trained on approximately 680,000 hours of multilingual, weakly supervised speech data, enabling strong generalization across various speech-related tasks. Whisper is released in multiple variants with different parameter scales. Among them, Whisper-large~\cite{ma2024extending}, with approximately 1.543 billion parameters, achieves leading performance on a range of ASR benchmarks and exhibits strong robustness in low-resource languages, noisy environments, and multi-speaker scenarios.

However, since Whisper is pretrained almost entirely on normal speech, its performance degrades significantly when applied to dysarthric speech. To address this limitation, we employ the proposed DARS-TTS system to generate large-scale dysarthria-prosody synthetic data for Whisper fine-tuning. 
We explore two adaptation strategies: full-parameter fine-tuning and parameter-efficient Low-Rank Adaptation (LoRA)~\cite{liu2024exploration}, to facilitate effective adaptation to dysarthric speech.

\section{PROPOSED METHOD}

Building upon Matcha-TTS, we introduce rhythm and style adaptation mechanisms specifically designed for dysarthric speech. 
The following sections elaborate on each of these mechanisms in detail.

\subsection{Multi-Stage Rhythm Predictor Optimized by Dysarthria-Guided Contrastive Preference
}
The duration predictor in Matcha-TTS is used to estimate durations from the input text. However, since the text is typically preprocessed based on normal speech, its pause patterns often differ significantly from those in dysarthric speech. 
To address this, we introduce a dysarthric speech rhythm predictor (blue box in Fig.~\ref{framework_v1}), which integrates multi-stage prediction and CPO to precisely control the distinctive rhythmic patterns.

\subsubsection{Multi-Stage Rhythm Prediction}
The multi-stage rhythm predictor models abnormal rhythmic patterns in pathological speech through a sequential processing pipeline. First, the raw phoneme sequence is encoded by the phoneme encoder without pause information. A pause predictor then classifies potential pause types between phonemes and inserts corresponding pause embeddings into the sequence. The augmented sequence is re-encoded by the pause-augmented encoder to obtain contextual representations enriched with pause cues. Finally, these representations are fed into a duration predictor to estimate phoneme-level durations. This cascaded design integrates pauses explicitly into duration modeling, enabling more accurate reconstruction of dysarthric speech prosody.

Let the high-level phonemic representation sequence from the phoneme encoder be  $\mathbf{E}$ = $[\mathbf{e}_{1},\mathbf{e}_{2},\cdots,\mathbf{e}_{N}]$.
The pause predictor $f^{s}_\theta$ processes $\mathbf{E}$ to output pause class distributions $\hat{\mathbf{S}} = [\hat{\mathbf{s}}_{1}, \hat{\mathbf{s}}_{2}, \cdots, \hat{\mathbf{s}}_{N}]$.
Ground-truth labels $\mathbf{S} = [s_1, s_2, \cdots, s_N]$ are derived from alignments obtained using the Montreal Forced Aligner (MFA)~\cite{mcauliffe2017montreal}. Specifically, pauses are grouped into 
${K}$ discrete categories according to their durations: class $0$ indicates no pause, and classes $1$ through ${K-1}$ represent increasingly longer pauses.
We use the cross-entropy loss to guide pause prediction:
\begin{small}
\begin{align}
\mathcal{L}_{s} &= -\frac{1}{N} \sum_{i=1}^{N} \sum_{k=0}^{K-1} \mathbb{I}(s_i = k) \log \hat{s}_{i,k}
\end{align}
\end{small}
where $\hat{s}_{i,k}$ denotes the predicted probability that the $i$-th position corresponds to pause class $k$.

Following pause prediction, pause embeddings (corresponding to $\hat{\mathbf{S}}$) are inserted into $\mathbf{E}$, creating an augmented sequence $\mathbf{E_{\text{aug}}}$. The pause-augmented encoder processes $\mathbf{E_{\text{aug}}}$ to obtain contextual representations enriched with pause information. These representations are then fed into the duration predictor $f^{t}_\theta$ to estimate durations.
We train the duration predictor using mean squared error (MSE) on log-scale durations:
\begin{small}
\begin{align}
\mathcal{L}_{d} &= \frac{1}{N^{\prime}} \sum_{i=1}^{N^{\prime}} (\log d_{i} - \log \hat{d}_{i})^2
\end{align}
\end{small}
\noindent where $\hat{d}_{i}$ is the predicted duration of the $i$-th element, $d_{i}$ is the reference duration obtained via the MAS alignment method~\cite{mehta2024matcha,popov2021grad}, and $N^{\prime} \geq N$ denotes the sequence length after pause insertion.

\subsubsection{Dysarthria-Guided Contrastive Preference Optimization}
Considering the complexity and variability of dysarthric speech patterns, we introduce a dysarthria-guided contrastive preference learning mechanism built upon the multi-stage rhythm prediction module.
It encourages the synthesized rhythmic style to better match the distribution of real dysarthric speech, while deviating from that of normal speech. 
We formulate the contrastive preference loss function $\mathcal{L}_{cp}$ as follows:
\vspace{-0.2cm}
\begin{small}
\begin{align}
\mathcal{L}_{cp} &= \frac{1}{N}\sum_{i=1}^{N} w_{i} \cdot \max\left(0,\, \mathrm{D}(\hat{d}_{i},d_{i}) -\mathrm{D}(\hat{d}_{i},d_{i}^{n}) + m \right) \\
w_i &= \begin{cases} 
\alpha \cdot \hat{s}_{i, s_i} & \text{if } s_i > 0 \\ 
\beta \cdot \hat{s}_{i, 0} & \text{if } s_i = 0 
\end{cases}
\end{align}
\end{small} 
\noindent where $\mathrm{D}(\hat{d}_{i},{d}_{i})$ represents the distance between the predicted duration $\hat{d}_{i}$ and the ground-truth dysarthric duration $d_{i}$, while $\mathrm{D}(\hat{d}_{i},d_{i}^{n})$ is the distance between $\hat{d}_{i}$ and the duration of normal speech ${d}_{i}^{n}$. In our experiments, this distance is measured by the absolute difference. The margin $m$ is a predefined threshold, typically set to $0.75$. 
The contrastive loss $\mathcal{L}_{cp}$ enforces that the predicted duration should be closer to the dysarthric ground truth than to the normal-speech counterpart by at least $m$.
The dynamic weight $w_i$ is defined based on the model's predicted class probabilities. For positions labeled as pauses ($s_i > 0$), the weight is proportional to $\hat{s}_{i, s_i}$, the predicted probability of the true pause class. 
For non-pause positions ($s_i = 0$), the weight depends on $\hat{s}_{i, 0}$, the probability assigned to the non-pause class.
Parameters $\alpha$ and $\beta$ 
control the relative weights assigned to pause and non-pause positions in the loss function.
We empirically set $\alpha \geq \beta$ to emphasize accurate pause modeling, which is critical for capturing the atypical rhythmic patterns characteristic of dysarthric speech.

\subsection{Dysarthria-aware Acoustic Conditional Flow Matching}

Matcha-TTS predicts the flow matching vector field solely based on the conditional mean values $\bm{\mu}$, derived from the phoneme sequence $\mathbf{E}$ and speaker identity $\mathbf{v}$. However, in the context of dysarthric speech synthesis, such a design may overlook rich pathological prosodic patterns embedded in the acoustic features. To address this limitation, we propose a dysarthria-aware acoustic conditional flow matching mechanism, which introduces additional global and local acoustic style vectors, $\mathcal{A}_g$ and $\mathcal{A}_l$, which are fused with content-related conditions to jointly modulate the flow matching process.

Specifically, we extract global style tokens (GSTs)~\cite{pmlr-v80-wang18h} from the reference mel-spectrogram as the global prosodic representation $\mathcal{A}_g$. 
Concurrently, we model frame-level style variations through a local style encoder adapted from~\cite{NEURIPS2022_4730d10b}, yielding $\mathcal{A}_l$, which employs a vector quantization bottleneck to extract style representations from reference mel-spectrograms and aligns them with the pause-augmented encoder's output hidden states $\mathcal{H}_c$ via attention.
The aligned local style features $\mathcal{A}_l$ are then fused with $\mathcal{H}_c$ to condition the generation of the conditional mean values $\bm{\mu}$.
The complete vector prediction network is defined as follows:
\begin{small}
\begin{align}
\bm{\mu} &= \text{Encoder}(\mathbf{E}, \mathcal{A}_g, \mathcal{A}_l)  \\
\mathbf{v}_{t}(\mathbf{x} \mid \bm{\mu}; \bm{\theta}) &= \text{Decoder}(\mathbf{x}, \bm{\mu}, t)
\end{align}
\end{small}
\noindent where $\text{Decoder}(\cdot)$ is a U-Net-based vector field prediction network, $\bm{\theta}$ denotes the model parameters, $\mathbf{x}$ is the latent variable, and $t \in [0, 1]$ represents the flow matching step.

Ultimately, our model adopts the conditional flow matching objective $\mathcal{L}_{\text{CFM}}$ from Matcha-TTS~\cite{mehta2024matcha}, which aims to learn a vector field that maps samples from a simple prior distribution to dysarthric acoustic features.  
Beyond content-related information, our approach further incorporates dysarthric acoustic style representations into the vector field prediction process, thereby enhancing the model's ability to reconstruct typical acoustic abnormalities found in pathological speech.


\section{EXPERIMENTS}

\subsection{Experimental Setup}
The TORGO database is employed as the dysarthric speech dataset~\cite{rudzicz2012torgo}.
It contains recordings from 8 speakers with dysarthria and 7 healthy control speakers. The 8 dysarthric speakers have been diagnosed with either Amyotrophic Lateral Sclerosis (ALS) or Cerebral Palsy (CP). As shown in Table~\ref{experiment_1}, these speakers are categorized into four severity levels based on their speech impairments: Severe, Mod.-Sev., Moderate, and Mild.
In the speaker descriptions, "F" and "M" indicate female and male genders, respectively, while the accompanying numbers denote participant IDs within the dataset.

\begin{table}[H]
\centering
\caption{Dysarthria severity for the TORGO database.}
\setlength{\tabcolsep}{1.0 mm}{
\fontsize{8}{12}\selectfont
\begin{tabular}{ccccccccc}
\toprule

                                                                & \multicolumn{4}{c}{Severe} & Mod.-Sev. & Moderate & \multicolumn{2}{c}{Mild} \\ \hline
Participant                                                     & F01   & M01  & M02  & M04  & M05      & F03      & F04         & M03        \\ \hline
\begin{tabular}[c]{@{}c@{}}Number of \\ Utterances\end{tabular} & 228   & 739  & 772  & 659  & 610      & 1097     & 675         & 806      \\

\bottomrule
\end{tabular}}
\label{experiment_1}
\end{table}

Given that the TORGO dataset is resource-limited, we investigate training strategies for the Matcha-TTS model to minimize data usage as much as possible:

\begin{itemize}
\item All-Speaker (ASp): A single TTS model is trained using the combined training data from all dysarthric speakers.
\item Single-Speaker (SSp): 
Eight individual models are trained separately for each dysarthric speaker (F01$\sim$M03), using only their corresponding speech data.
\item Dysarthria-Severity-Group Speaker (DSpG): Four TTS models are trained independently on data grouped by severity levels of dysarthria.
\end{itemize}

For the Matcha-TTS model architecture and training strategies, we adopt the MAT-10 configuration as described in~\cite{mehta2024matcha}. However, we expand the parameter size of the text encoder, which now comprises 8 layers with 4 attention heads and 1024 filter channels per layer.
In our proposed DARS framework, the phoneme encoder and the pause-augmented encoder follow architectures similar to those of the original Matcha-TTS encoder. Specifically, the phoneme encoder is a smaller variant consisting of only 2 layers with 4 attention heads and 512 filter channels per layer, whereas the pause-augmented encoder is larger, matching the previously described text encoder in scale.
The pause and duration predictors share identical network structures.
Further details regarding the GST encoder can be found in~\cite{pmlr-v80-wang18h}, while the implementation of the frame-level local style encoder follows~\cite{NEURIPS2022_4730d10b}.
Prior to training the DARS model, we employ an MFA model~\cite{mcauliffe2017montreal} trained on the full TORGO dataset to extract pause labels.
In CPO, the duration predictor for normal speech is pre-trained on the LibriSpeech English dataset~\cite{panayotov2015librispeech} to obtain the reference duration corresponding to the input text.

For dysarthric ASR, we fine-tune the Whisper-Large model on the synthesized speech data and evaluate its performance on the original data. Since the TORGO dataset does not provide predefined splits for training, validation, and evaluation, we employ a multi-sampling strategy, selecting 10\%, 20\%, and 30\% of the data as evaluation sets. These subsets are chosen to ensure minimal performance variation across the three splits when evaluated with the publicly available Whisper-Large model~\cite{cao2012whisper}. 
The remaining data is then divided into training and validation sets with a 9:1 ratio, which are also used for training the Matcha-TTS model. 
Subsequently, the trained Matcha-TTS is used to synthesize the training data, which is employed for Whisper-Large adaptation.

We use Mean Cepstral Distortion (MCD)~\cite{kominek2008synthesizer, leung2024training} to evaluate the similarity between the synthesized Mel-spectrogram and the ground-truth Mel-spectrogram.
In speech synthesis evaluations, MCD has been shown to correlate with subjective evaluation results. 
For ASR performance evaluation, we use the Word Error Rate (WER)~\cite{cao2012whisper} as the primary metric.
Additionally, an Overall WER is calculated by averaging the WER scores across individual speakers.

\subsection{Experimental Results}
~\textbf{Evaluation Set Stability Across Sampling Ratios.} Table~\ref{experiment_2} presents the performance of the open-source Whisper-Large model on TORGO test sets sampled at different proportions.
These evaluation sets (E1--E3)
were sampled and recognized multiple times, demonstrating robust stability across sampling ratios.
Based on this consistency, we select the 20\% evaluation set for all subsequent experiments.
The remaining 80\% of the data serves as training and validation data for both speech synthesis and recognition tasks.

\begin{table}[!t]
\centering
\caption{WER Comparison on Torgo Evaluation Sets by Sentence Count.}
\setlength{\tabcolsep}{1.0 mm}{
\fontsize{8}{12}\selectfont
\begin{tabular}{ccccccc}
\toprule

\multirow{2}{*}{ID} & \multirow{2}{*}{Whisper-Large} & \multicolumn{5}{c}{WER (\%)}                                                          \\ \cline{3-7} 
                    &                                & Severe          & Mod.-Sev.        & Moderate       & Mild           & Overall        \\ \hline
E1                  & 10\%                            & 125.67          & 182.53          & 32.78          & 13.91          & 83.21          \\
E2                  & 20\%                            & \textbf{123.36} & \textbf{184.14} & \textbf{34.69} & \textbf{12.71} & \textbf{83.19} \\
E3                  & 30\%                            & 121.91          & 186.25          & 35.47          & 11.98          & 83.24      \\

\bottomrule
\end{tabular}}
\label{experiment_2}
\end{table}

~\textbf{Impact of Modeling Mechanisms on Synthesis Quality.} Table~\ref{experiment_3} presents the MCD results on the validation set for synthesized speech generated by TTS models under three training strategies: ASp, DSpG, and SSp. E4 corresponds to the results from Grad-TTS. E5 serves as the baseline system using the original Matcha-TTS. E6 incorporates multi-stage rhythm prediction modules into E5. E7 and E8 further integrate the CPO strategy based on E6. E9 introduces acoustic style vectors into the CPO-enhanced models.




\begin{table}[!t]
\centering
\caption{MCD for TTS Models Trained on TORGO Data.} 
\setlength{\tabcolsep}{1.5 mm}{
\fontsize{8}{12}\selectfont
\begin{tabular}{c l c c c c c}
\toprule
ID  & \multicolumn{1}{c}{Models} & $\alpha$ & $\beta$ & ASp    & DSpG   & SSp    \\ 
\midrule
E4  & Grad-TTS~\cite{leung2024training}    & --      & --      & 6.61   & 6.71   & 6.81   \\
\midrule
E5  & MATCHA-TTS (baseline)                & --      & --      & 6.25   & 6.43   & 6.64   \\ 
\midrule
E6  & DARS (E5 w/ rhythm)                  & --      & --      & 6.09   & 6.32   & 6.57   \\ 
\midrule
E7  & \multirow{2}{*}{DARS w/ CPO}      & 0.5     & 0.5     & 5.83   & 6.07   & 6.21   \\
E8  &                                      & 0.7     & 0.3     & 5.72   & 5.93   & 6.08   \\ 
\midrule
E9 & DARS w/ CPO + style      & 0.7     & 0.3     & \textbf{4.29} & \textbf{4.46} & \textbf{4.61} \\
\bottomrule
\end{tabular}}
\label{experiment_3}
\end{table}
We observe that the ASp training strategy achieves the best performance, which can be attributed to its ability to utilize a larger amount of training data. Incorporating pause and duration modeling, together with the CPO strategy, significantly reduces the acoustic discrepancy between synthesized and real dysarthric speech. Furthermore, the integration of acoustic style vectors further improves synthesis quality by providing speaker-specific acoustic guidance.
In the following experiments, we analyze the synthesized data based on the three training strategies derived from E9.

~\textbf{Effectiveness of Synthesized Speech in Enhancing ASR.} 
Table~\ref{experiment_4} compares the recognition performance of ASR systems trained on dysarthric speech synthesized by TTS models using different training strategies versus systems trained on real speech. 
We primarily adopt two adaptation strategies for E9-synthesized data: full-parameter fine-tuning and LoRA-based fine-tuning. 
It can be observed that full-parameter fine-tuning outperforms LoRA. 
This performance gap likely stems from the Whisper model's original pre-training on non-dysarthric speech.
As a result, LoRA’s localized parameter adaptation shows limited modeling capacity when applied to dysarthric speech, which exhibits significant distributional differences from typical speech patterns.

\begin{table}[!t]
\centering
\caption{WER Performance Comparison Across Data Types and Training Strategies Using the Whisper-Large Model.}
\setlength{\tabcolsep}{1.0 mm}{
\fontsize{7.5}{12}\selectfont
\begin{tabular}{ccccccccc}
\toprule

\multirow{2}{*}{ID} & \multirow{2}{*}{Data Type} & \multicolumn{2}{c}{training strategy} & \multicolumn{5}{c}{WER (\%)}                                                     \\ \cline{3-9} 
                    &                            & FT               & LORA               & Severe         & Mod.-Sev.       & Moderate      & Mild          & Overall       \\ \hline
E2                  & -                          &                  &                    & 123.36         & 184.14         & 34.69         & 12.71         & 83.19         \\ \hline
E10                 & \multirow{2}{*}{RealData}  & \(\checkmark\)                &                    & 18.63          & 11.88          & 2.93          & 2.43          & 8.85          \\
E11                 &                            &                  & \(\checkmark\)                  & 22.35          & 13.66          & 3.24          & 2.54          & 10.09         \\ \hline
E12                 & \multirow{2}{*}{ASp}       & \(\checkmark\)                &                    & \textbf{18.64} & \textbf{11.88} & \textbf{2.94} & \textbf{2.44} & \textbf{8.87} \\
E13                 &                            &                  & \(\checkmark\)                  & 22.37          & 13.67          & 3.24          & 2.56          & 10.12         \\ \hline
E14                 & \multirow{2}{*}{DSpG}      & \(\checkmark\)                &                    & 21.06          & 13.07          & 3.15          & 2.50          & 9.96          \\
E15                 &                            &                  & \(\checkmark\)                  & 24.23          & 15.45          & 3.21          & 2.54          & 11.32         \\ \hline
E16                 & \multirow{2}{*}{SSp}       & \(\checkmark\)                &                    & 22.18          & 13.55          & 3.21          & 2.54          & 10.31         \\
E17                 &                            &                  & \(\checkmark\)                  & 26.10          & 16.64          & 3.30          & 2.61          & 12.19     \\

\bottomrule
\end{tabular}}
\label{experiment_4}
\end{table}
It can be observed that the ASp training strategy (E12) generates dysarthric speech that closely matches real speech in acoustic characteristics and demonstrates strong generalization capability.
ASR models trained on E12 achieve recognition performance comparable to those trained on real speech (E10) across all dysarthria severity levels.
Moreover, compared to the baseline model (E2), E12 achieves an 89.33\% relative reduction in Overall WER.
Notably, greater performance gains are observed for more severe impairment categories, highlighting the enhanced effectiveness and adaptability of DARS in addressing significant speech impairments.

~\textbf{Comparison with SOTA Systems and DARS Model.} 
Table~\ref{experiment_5} presents a comparison of recognition performance across different severity levels on the TORGO dataset between DARS and state-of-the-art (SOTA) systems. In E18~\cite{soleymanpour2022synthesizing}, the TTS model is FastSpeech2~\cite{ren2020fastspeech} and the ASR model is a DNN-HMM~\cite{li2013hybrid} system. In E19~\cite{leung2024training}, the TTS model is Grad-TTS~\cite{popov2021grad} and the ASR model is Whisper.

\begin{table}[!t]
\centering
\caption{WER Comparison with SOTA Systems.}
\setlength{\tabcolsep}{1.0 mm}{
\fontsize{8}{12}\selectfont
\begin{tabular}{ccccccccc}
\toprule

\multirow{2}{*}{ID} & \multirow{2}{*}{DataSet}                                          & \multicolumn{5}{c}{WER (\%)}                                                    \\ \cline{3-7} 
                    &                                                                   & Severe         & Mod.-Sev.      & Moderate      & Mild          & Overall       \\ \hline
E18~\cite{soleymanpour2022synthesizing}              & TORGO                                                             & 55.88          & 49.6          & 36.8          & 12.6          & 39.2          \\ \hline
E19~\cite{leung2024training}              & TORGO                                                             & 23.3           & 13.98         & 3.27          & 2.57          & 16.93         \\ \hline
E12                 & TORGO                                                             & 18.64          & 11.88         & 2.94          & 2.44          & 8.87          \\ \hline
E20                 & \begin{tabular}[c]{@{}c@{}}TORGO+\\ LibriSpeech Text\end{tabular} & \textbf{13.98} & \textbf{9.79} & \textbf{2.62} & \textbf{2.31} & \textbf{7.75}     \\

\bottomrule
\end{tabular}}
\label{experiment_5}
\end{table}
To evaluate DARS's generalization ability,
we perform dysarthric speech synthesis using text from the LibriSpeech dataset, with reference speech randomly selected from TORGO. After adapting the ASR model using the augmented dysarthric speech (E20), we achieve an Overall WER of 7.75\%, representing a 54.22\% relative WER reduction over E19.

\section{CONCLUSIONS}
This study demonstrates that fine-tuning the Whisper model with DARS-synthesized speech achieves significant WER reductions across all dysarthria severity levels, with a relative reduction of 93.54\% in Mod.-Sev. cases. Compared to advanced systems combining Grad-TTS and Whisper, our approach delivers a 54.22\% relative WER reduction, demonstrating superior prosodic adaptation and generalization capabilities. Cross-corpus synthesis experiments using LibriSpeech text further confirm the robustness of our method in out-of-domain scenarios. These results indicate that joint modeling of pathological rhythm and acoustic styles not only enhances the realism and controllability of synthetic speech, but also significantly improves the effectiveness and generalizability of data augmentation for dysarthric speech recognition.










\printbibliography

\end{document}